\documentclass[aps,twocolumn,groupedaddress,showpacs]{revtex4}
\usepackage[dvips]{graphicx}
\usepackage[]{caption}
\usepackage{amsmath}
\usepackage{amssymb}
\def\Dsl{\hbox{/\kern-.6700em\it D}} % D slash
\def\dsl{\hbox{/\kern-.5300em$\partial$}}

\def\eqa{\begin{eqnarray}}
\def\eeqa{\end{eqnarray}}
\def\eq{\begin{equation}}
\def\eeq{\end{equation}}
\def\be{\begin{equation}}
\def\ee{\end{equation}}
\def\bea{\begin{eqnarray}}
\def\eea{\end{eqnarray}}

\begin{document}
\bibliographystyle{prsty}
\title{Brane gas-driven bulk expansion as a precursor stage to brane
inflation}
\author{Natalia Shuhmaher and Robert Brandenberger}
\affiliation{Dept. of Physics, McGill University, 3600 University Street,
Montr\'eal QC, H3A 2T8, Canada}
\date{\today}
\begin{abstract}

We propose a new way of obtaining slow-roll inflation in the context
of higher dimensional models motivated by string and M theory. In
our model, all extra spatial dimensions are orbifolded. The initial
conditions are taken to be a hot dense
bulk brane gas which drives an initial phase of isotropic bulk
expansion. This phase ends when a weak potential between the orbifold
fixed planes begins to dominate. For a wide class of potentials,
a period during which the bulk dimensions decrease sufficiently
slowly to lead to slow-roll inflation of the three dimensions parallel
to the orbifold fixed planes will result.
Once the separation between the orbifold
fixed planes becomes of the string scale, a repulsive potential due
to string effects takes over and leads to a stabilization of the
radion modes. The conversion of bulk branes into
radiation during the phase of bulk contraction leads to reheating.
\end{abstract}
\maketitle

%%%%%%%%%%%%%%%%%%%%%%%%
{\bf 1. Introduction}:
The Inflationary Universe scenario \cite{Guth} (see also
\cite{Sato,Brout,Starob})  has been extremely successful
phenomenologically. It has provided a solution to some of the key
problems of standard cosmology, namely the horizon and flatness
problems, and yielded a mechanism for producing
primordial cosmological perturbations using causal physics, a
mechanism which predicted \cite{ChibMukh,Lukash} (see also
\cite{Press,Sato}) an almost scale-invariant spectrum of adiabatic
cosmological fluctuations, a prediction confirmed more than a decade
later to high precision by cosmic microwave background anisotropy experiments
\cite{COBE,Boomerang,Maxima,WMAP}.

However, it has proven difficult to find convincing realizations of
inflation in the context of quantum field theory models of matter in four
space-time dimensions. It is usually assumed that the quasi-constant
potential energy of a slowly rolling scalar field (the so-called
``inflaton'') leads to the accelerated
expansion which inflation requires. The Standard Model of
particle physics, however, does not contain a scalar field
whose dynamics leads to slow-rolling. In single field models with a
renormalizable potential, field values larger than $m_{pl}$ (the
four-dimensional Planck mass) are required in order to obtain
slow-rolling as a local attractor in the phase space of homogeneous
solutions to the scalar field equations of motion \cite{Kung}.

Superstring theory and M-theory, on the other hand, contain a lot
of degrees of freedom which at the level of the four space-time dimensional
effective field theory are described by scalar fields. Supersymmetry
ensures that some of these fields (the so-called ``moduli fields''
are sufficiently weakly coupled to provide potential candidates to
be an inflaton.

In the context of brane world cosmology \cite{ADD,RS}, an
appealing possibility is that the separation between
a brane and an antibrane \cite{DT,Shafi} can serve as the inflaton. A
problem with the proposed constructions, which were all in the
context of a static bulk, was that the bulk size was generically
too small to allow for the large values of the inflaton field required
to generate inflation. This problem was addressed in \cite{Cliff,Cliff2}.
Another possibility is to have topological brane inflation \cite{Stephon},
but this construction also requires special parameters in order to obtain
a wide enough brane.

The constructions mentioned in the previous paragraph were all done
in the context of phenomenological field theoretical models inspired
by string theory. After the discovery that flux constructions can lead
to a stabilization mechanism for most moduli fields of string theory
\cite{Sethi,GKP}, a lot of attention (beginning with \cite{KKLT,Keshav})
was focused on how to obtain inflationary models in the context
of flux compactifications (see \cite{stringinflation} for reviews
and comprehensive lists of references). These constructions are, once
again, in the context of static bulk configurations, and have to
assume very special configurations (special configurations of branes
and special flux choices).

In this paper, we present a new model of brane inflation. In contrast
to previous constructions, the dynamics of the bulk dimensions is
essential to our model. Also, in contrast to previous constructions,
we start with initial conditions which we consider to be very natural,
namely a hot brane gas in the context of an initial universe in which
all spatial dimensions are democratically small (of the string scale),
similar to what is assumed in ``string gas cosmology'' \cite{BV}
and its brane generalizations \cite{ABE,Mahbub,Mairi,Lisa}). The hot
brane gas leads to an initial phase of isotropic bulk expansion
(Phase 1 of our cosmology). During this phase, the bulk energy density
decreases.

The isotropy of space is explicitly broken by our
assumption that the extra spatial dimensions are orbifolded. This
leads to the existence of orbifold fixed planes.
We assume the existence of a weak attractive potential
between the orbifold fixed planes \footnote{Note that in
terms of having inflation driven by the potential between orbifold
fixed planes, our setup is similar to that of \cite{Cliff2}.}.
Eventually, the associated
potential energy will begin to dominate the dynamics and will lead
to a contraction of the dimensions perpendicular to the orbifold
fixed planes (Phase 2). We will consider a potential of the form
\be \label{pot}
V(r) \, = \, - \mu {1 \over r^n} \, ,
\ee
where $r$ is the separation of the orbifolds, and $n$ is an exponent
which we will fix later. Such a potential could emerge from charges
on branes pinned to the orbifold fixed planes. We will show that such a potential
can lead to slow-roll inflation. The inflationary slow-roll
parameters are set by the coefficient
$\mu \equiv \Lambda^{4 + d - n}$ (where $\Lambda$ has dimensions of
energy) which characterizes the strength of the potential.
The requirement of a sufficient number of e-foldings to solve the
cosmological problems of standard cosmology \cite{Guth} sets an
upper bound on $\Lambda$.

Once $r$ decreases to the string scale, a repulsive potential
created by stringy effects will take over. The competition
of the repulsive short range force and the attractive long range
force, together with the continued expansion of space parallel
to the orbifold fixed planes, will lead to a stabilization of
$r$. This stabilization scenario is an application of the mechanism
of radion stabilization which has recently been studied extensively
in the context of string gas cosmology \cite{Patil1,Patil2,Watson2} (see
\cite{RHBrev3} for a short review). Either during Phase 2, or once
the separation of the orbifold fixed planes has decreased to the
string scale, the bulk branes annihilate and decay into radiation.
This leads to a smooth transition into the radiation phase of
standard cosmology.

Note that a very similar setup was used recently \cite{size} to
construct a non-inflationary solution to the entropy and horizon
problems of standard cosmology. In \cite{size}, we assumed that
the inter-brane potential was confining, a potential of the type
that could be generated by non-perturbative effects. Here, we
take the potential (\ref{pot}) which could come from string exchange
between branes \cite{origin}.

%%%%%%%%%%%%%%%%%%%%
{\bf 2. The Model}:
Our starting point is a topology of space in which all but three
spatial dimensions are orbifolded, and the three dimensions
corresponding to our presently observed space are
toroidal. Specifically, the space-time manifold is
\be \label{orbi}
{\cal M} \, = \, {\cal R} \times T^3 \times T^d / Z_2
\, ,
\ee
where $T^3$ stands for the three-dimensional torus, and $d$ is the
number of extra spatial dimensions, which we will take to be
either $d = 6$ in the case of models coming from superstring theory,
or $d = 7$ in the case of models motivated by M-theory. We will assume
that there is a weak force between the orbifold fixed planes given by
the potential (\ref{pot})
\footnote{It may be necessary to have branes pinned to the orbifold fixed
planes in order to induce such a potential. Our approach, at this stage,
is purely phenomenological, and we simply postulate the existence of a
potential with the required properties}.

As our initial conditions, we take the bulk to be filled with a gas of
p-branes, as in our previous work \cite{size}. In the context of Type IIB
superstring theory we will have D-branes with $p = 3$,
in the case of heterotic string theory or taking the starting point to
be M-theory, we have Neveu-Schwarz 5-branes ($p = 5$).

Assuming that the universe starts out small and hot, it is reasonable
to assume that the energy density in the brane gas will initially be
many orders of magnitude larger than the potential energy density
generated by the force between the orbifold fixed planes. Thus,
initially our universe will be expanding isotropically. As shown in
\cite{size}, this expansion is non-inflationary.
During this phase (Phase 1), the bulk energy density will decrease. Hence,
eventually the inter-orbifold potential will begin to dominate. At
this point, the cosmological evolution will cease to be isotropic: the
directions parallel to the orbifold fixed planes will continue to
expand while the perpendicular dimensions begin to contract: this
marks the beginning of Phase 2. In the following we will show that
for a wide class of potentials, the expansion of our dimensions
will be inflationary.

The metric in the non-isotropic phase (and in the absence of
spatial curvature) is given by
\be
ds^2 \, = \, dt^2 - a(t)^2 d{\bf x}^2 - b(t)^2 d{\bf y}^2 \, ,
\ee
where ${\bf x}$ denote the three coordinates parallel to the boundary
planes and ${\bf y}$ denote the coordinates of the perpendicular
directions. Hence, the radius $r$ of the dimensions perpendicular
to the orbifold fixed planes is given by
%\be
$ r(t) \, = \, l_s b(t) \, .$
%\ee

We will analyse the evolution during Phase 2 using a four-dimensional
effective field theory, where we replace the radion $b(t)$ by a scalar
field $\varphi(t)$. In order that $\varphi$ be canonically normalized
when starting from the higher dimensional action of General Relativity,
$\varphi$ and $b$ must be related via (see e.g. \cite{BattWat}, Appendix A
for a review)
\be \label{rel}
\varphi \, = \, {\sqrt{2d}} m_{pl} {\rm log}(b) \, .
\ee
If the bulk
size starts out at the string scale, then $b(t_b) = 1$, where $t_b$
is the initial time. With these normalizations, $\varphi = 0$
corresponds to string separation between the branes. The
dimensional reduction of the higher dimensional gravitational
action to the four space-time dimensional Einstein frame action
yields the following effective potential for $\varphi$ \cite{BattWat}
\be
V_{eff}(\varphi) \, = \, l_s^d b(\varphi)^{-d} V(r(\varphi)) \, .
\ee
Note that the dilaton is assumed to be fixed, and the dilaton-dependence of
the potential is neglected.
From the potential (\ref{pot}) and inserting the relation (\ref{rel})
we obtain
\bea \label{pot2}
V_{eff}(\varphi) \, &=& \, - \Lambda^{4 + d - n}
l_s^{d - n} e^{- {{d + n} \over {\sqrt{2d}}} {{\varphi} \over {m_{pl}}}}
\nonumber \\
&=& \, - \Lambda^{4 + d - n} l_s^{d - n}
e^{- {\tilde \alpha} {{\varphi} \over {m_{pl}}}}
\, ,
\eea
where we have defined a constant
${\tilde \alpha} \equiv  (d + n) / {\sqrt{2d}}$.
As an example, for $n = 1$ and $d=6$ we obtain ${\tilde \alpha} = 2.02$.

To ensure
vanishing of the four-dimensional cosmological constant today,
we must add a positive constant $V_0$ to the effective potential
(\ref{pot2}). If the stabilization radius of the extra dimensions
is the string scale $l_s$, then $V_0$ is given by
\be
V_0 \, = \, \Lambda^{4 + d - n} l_s^{d - n} \, .
\ee

{F}rom the form of the potential, it should be expected that a
period of slow-roll inflation is possible as long as the initial
value of $\varphi$ at the beginning of Phase 2 is larger than
$m_{pl}$. The special feature of our scenario (and the major
advantage compared to previous versions of brane inflation), is
that such large values of $\varphi$ dynamically emerge and do not
have to be inserted as ad hoc initial conditions.

In our scenario, inflation has a graceful exit. Once the orbifold
fixed planes reach a microscopic separation, Kaluza-Klein momentum
modes of strings (e.g. the momenta of the massless states produced
at enhanced symmetry points) produce a repulsive potential
which scales as $b^{-2}$~\cite{Watson,Beauty} and hence on short distances
overwhelms the large-distance attractive potential (provided $n <
2$). The interplay
between this repulsive potential which dominates at small separations and
the attractive potential which dominates at large distances, coupled to
the expansion of the three dimensions parallel to the orbifold fixed planes,
will lead to the stabilization of the radion modes at a specific radius
(presumably related to the string scale). In the context of heterotic string
theory, we could use the string states which are massless at the
self-dual radius to obtain stabilization of the radion modes at the
self-dual radius \cite{Patil1,Patil2} (see also \cite{Watson2}). These
modes would also ensure dynamical shape moduli stabilization
\cite{Edna}. The branes decay into radiation either during or at the
end of Phase 2. This brane decay is the main source of reheating of
our three dimensional space.
We denote the time of radion stabilization and reheating
by $t_R$. After reheating, our three spatial dimensions emerge in the
radiation phase of standard cosmology.

%%%%%%%%%%%%%%%%%%%%%%%%%%%%%%%%%%%%
{\bf 3. The Phase of Bulk Expansion}:
The phase of isotropic bulk expansion ($a(t) = b(t)$) proceeds as
discussed in \cite{size}. The equation of motion for $a(t)$ is
\be \label{bulk}
{{{\ddot a}} \over a} + (2 + d)\bigl({{{\dot a}} \over a} \bigr)^2
\, = \, {{8 \pi G} \over {3 + d - 1}} \bigl[ \rho - P \bigr] \, ,
\ee
where $\rho$ is the energy density and $P$ denotes the pressure.
Making use of the equation of state $P = w \rho$, and inserting
the Einstein constraint equation
\be \label{constr}
\bigl((3 + d)^2 - 3 - d\bigr)H^2  \, = \, 16 \pi G \rho \, ,
\ee
where $H \equiv {\dot a}/a$, we obtain power law expansion
\be \label{scaling2}
a(t) \, \sim t^{\alpha} \,\,\, {\rm with} \,\,\,
\alpha \, = \, {2 \over {(3 + d)(1 + w)}} \, .
\ee
In the case of bulk energy dominated by the tension of $p$-branes, we have
\be
w \, = \, - {p \over {3 + d}} \, .
\ee
Thus, in the example motivated by perturbative Type IIB superstring theory,
($d = 6$ and $p = 3$) we obtain $\alpha = 1/3$.
Starting with heterotic string theory ($d = 6$ and $p = 5$) we obtain
$\alpha = 1/2$, and for M-theory ($d = 7$ and $p = 5$) we get
$\alpha = 2/5$. Thus, the phase of bulk expansion is non-inflationary.

%%%%%%%%%%%%%%%%%%%%%%%%%%%%%%%%%%%
{\bf 4. The Period of Inflation}:
The period of bulk expansion ends when the bulk potential
energy becomes equal to the bulk brane energy density.
The bulk energy density in Phase 1 scales as
\be
\rho_b(t) \, \sim \, a(t)^{- d - 3 + p}
\ee
Assuming that the initial bulk energy density is given by
the string scale, i.e.
\be
\rho_b(t_b) \, \sim \, l_s^{-4 - d} \,
\ee
where $t_b$ denotes the initial time, then the transition
between Phase 1 and Phase 2 takes place at a time $t_i$
given by
\be
b(t_i)^{-(d + 3 - p - n)} \, = \, \bigl(\Lambda l_s\bigl)^{d + 4 - n} \, .
\ee
The value of the radion $\varphi$ at this time is given by
\be
\varphi(t_i) \, = \, \sqrt{2d} m_{pl} {\rm log}(b(t_i)) \, .
\ee
Since $\varphi$ is canonically normalized, its
equation of motion is given by (see the form of the effective
potential of (\ref{pot2}))
\be \label{eom}
{\ddot \varphi} + 3H {\dot \varphi} \,
= \, - \Lambda^{4 + d - n} l_s^{d - n} {{\tilde \alpha} \over {m_{pl}}}
e^{- {\tilde \alpha} \varphi / m_{pl}} \, .
\ee
The slow-roll conditions are satisfied provided:
\be
\varphi \gg {m_{pl} \over {\tilde \alpha}} \log\Big\{1 + {\tilde \alpha}^2 \Big\} \, .
\ee
Thus, to get $N$ efolding of inflation, the initial value of $\varphi$ should exceed the bound \be \varphi(t_i) > {m_{pl} \over {\tilde \alpha}} \log\Big\{ {\tilde \alpha}^2 (N + 1) + 1 \Big\} \ee leading to the condition
\be \label{bound}
l_s \Lambda \, < \,
\Big[ {\tilde \alpha}^2 (N + 1) + 1 \Big]^{- {{d + 3 - p - n} \over {\sqrt{2d} {\tilde \alpha} (d + 4 - n)}}}
\ee
which allows $\Lambda$ of order of the string scale.

We conclude that, provided the bound (\ref{bound}) on the
energy scale $\Lambda$ is satisfied, Phase 2 will provide
a sufficient length of inflation of our three spatial dimensions,
inflation driven by the slow rolling of the modulus field.

%%%%%%%%%%%%%%%%%%%%%%%%%%%%%%%%%%%%
{\bf 5. Discussion and Conclusions}:
In this paper we have proposed a new way of obtaining
inflation in the context of theories with extra dimensions
and branes. We assume that our three spatial dimensions
are singled out by the orbifold construction of (\ref{orbi}),
and that there is a weak potential between branes pinned to
the orbifold fixed planes. We assume attractive potentials
such as could emerge if opposite charges were localized on
the two branes.

In our scenario, the universe begins small and hot, filled
with an isotropic gas of branes. This brane gas drives a
period of isotropic bulk inflation which continues until
the potential between the branes localized on the orbifold
fixed planes becomes dominant. We have shown that the
potential of the radion supports a period of slow-roll
inflation. The new feature of our model compared to other
models of brane inflation is that the large values of the
radion required to obtain sufficient inflation
are dynamically generated during the phase of bulk
expansion.

Inflation ends when the radion shrinks to string-scale values, the
bulk branes annihilate into radiation, and the radion becomes
stabilized by string gas effects.

An interesting lesson obtained by comparing our present
results with those of our preceding paper \cite{size} is that
the details of the potential between the orbifold fixed
planes is very important in determining the evolution of
our three spatial dimensions. For a sufficiently confining
potential, our three spatial dimensions never undergo a
period of accelerated expansion - but the period of bulk
expansion still enables us to solve the horizon and entropy
problems because the energy density of the brane gas
projected onto the orbifold fixed planes does not decrease.
The condition on the power $n$ appearing in the potential (\ref{pot})
in order to obtain inflation is $n < d + 3 - p$. This limit mirrors the requirement that orbifold fixed plane potential is diluted slower than energy density of p-branes in the bulk.  

{\bf Acknowledgments}:
We wish to thank Thorsten Battefeld and
Keshav Dasgupta for useful discussions and for
comments on our manuscript.
This work is supported by funds from McGill University,
by an NSERC Discovery Grant and by the Canada Research Chairs program. N.S. would like to acknowledge support from a Carl Reinhardt McGill Major Fellowship.

%\end{acknowledgments}

\end{document}